# DESIGN AND TESTING OF THE NEW MUON LAB CRYOGENIC SYSTEM AT FERMILAB

A. Martinez, A. L. Klebaner, J. C. Theilacker, B. D. DeGraff, and J. Leibfritz

Fermi National Accelerator Laboratory
Batavia, IL, 60510, USA

## ABSTRACT

Fermi National Accelerator Laboratory is constructing a superconducting 1.3 GHz cryomodule test facility located at the New Muon Lab building. The facility will be used for testing and validating cryomodule designs as well as support systems. For the initial phase of the project, a single Type III plus 1.3 GHz cryomodule will be cooled and tested using a single Tevatron style standalone refrigerator. Subsequent phases involve testing as many as two full RF units consisting of up to six 1.3 GHz cryomodules with the addition of a new cryogenic plant. The cryogenic infrastructure consists of the refrigerator system, cryogenic distribution system as well as an ambient temperature pumping system to achieve 2 K operations with supporting purification systems. A discussion of the available capacity for the various phases versus the proposed heat loads is included as well as commissioning results and testing schedule. This paper describes the plans, status and challenges of this initial phase of the New Muon Lab cryogenic system.

## INTRODUCTION

Fermi National Accelerator Laboratory is constructing a superconducting radio-frequency (RF) test facility located at the New Muon Lab building at Fermilab for testing 1.3 GHz cryomodules. The new test facility provides the necessary test bed to measure the performance of superconducting radio frequency cavities in a cryomodule (CM). A multi-phased approach is being taken at the facility beginning with the installation and testing of the Capture Cavity II (CC2), a single high-gradient 1.3 GHz 9-cell superconducting RF cavity installed in a cryomodule cryostat. This cavity was previously installed at the Fermilab MDB test facility where it operated for several years [1]. This initial phase will allow for commissioning of the cryogenic distribution system leading to the test cave as well as allow for full RF testing and commissioning.

The next phase involves testing of the CC2 cavity in combination with the first Type III plus cryomodule (CM1) consisting of eight 9-cell superconducting RF cavities within a cryomodule cryostat. This is the first Type III plus cryomodule in the U.S and was assembled by DESY, Fermilab and INFN.

Subsequent phases include expansion of the facility to accept two full RF units consisting of three cryomodules each as well as the addition of single cryomodule test stands. The expansion would require adding an underground extension to the tunnel as well as the addition of a new cryogenic plant to handle the additional heat loads. At this stage, the cryogenic system would be broken up into two sections. The existing Tevatron style refrigerator utilized for the initial phases would be used to operate the front end photo-injector consisting of the Capture Cavity 1 (CC1) and CC2 cryomodules with the new cryogenic plant supplying the cryomodule RF units. These latter test phases will also operate with an upgraded RF system as well as a newly installed beam line.

## CRYOGENIC INFRASTRUCTURE

### Refrigeration System

The Phase 1 New Muon Lab cryogenic system is designed to support operation of the CC2 and CM1. For this initial phase, a single Tevatron satellite refrigerator (South refrigerator) will be used. This unit is identical to the Tevatron style refrigerators used throughout Fermilab with many years of operation. The single Tevatron satellite refrigerator will operate in mixed mode – refrigeration and liquefaction. The refrigeration component of the plant capacity will be used to compensate for heat leak of the distribution system and 5 K shields of the CC2 cryostat and the CM1. Liquid nitrogen will be used for the high temperature shields, which is supplied by an outdoor 6000 gallon liquid nitrogen dewar through a transfer line that extends underneath the roadway and into the building. The refrigerator consists of wet and dry reciprocating expansion engines, a cold box and distribution valve box. The refrigerator is also equipped with a 500 liter helium dewar for liquid storage and to ease system upsets and recovery.

Helium gas is supplied to the cryogenic system from four high pressure (250 psig MAWP) storage tanks located outdoors in the Northwest corner of the facility. The tanks have a total capacity of 43,700 gallons. The tanks are connected to each other with electrically actuated valves for remote switching between tanks and connection to the cryogenic refrigerator located inside the building. Each tank has a connection at both ends for ease of initial purification should the tanks be contaminated. Two tube trailer fill stations are located near the tanks as well for initial gas transfer and as a supplemental helium supply.

The compressor system consists of two (North and South) two-stage 400 horsepower Mycom screw compressors, each capable of delivering 60 g/s of helium at 1.2 psig suction and 285 psig discharge. Both compressors are located at the Lab B building which is approximately a quarter of a mile from the New Muon Lab facility. The suction and discharge lines run underground and enter the NML high bay near the inventory control area. Each compressor was originally taken from the Tevatron compressor system along with the associated oil removal equipment and installed at Lab B.

Two permanently mounted in-line $LN_2$ adsorbers are used to remove contaminants from the gas stream during initial start-up. The adsorbers are installed in the refrigerator room and are supplied liquid nitrogen from the dewar supply bayonet can using u-tubes. A pneumatically actuated valve matrix is used for remote switching and isolation of adsorbers as needed. The flexibility of having two adsorbers allows for one to be operational while the second can be regenerated without interrupting purification operations.

**Cryogenic Distribution System**

The supply transfer line was constructed to connect the New Muon Lab cryogenic refrigerator system to the test cave. It consists of North and South bayonet cans located in the refrigerator room, each having bayonet connections to connect to each individual cryogenic circuit: single phase helium, two phase helium and nitrogen. Each bayonet can extends through the refrigerator room floor and connect to an expansion can suspended below the refrigerator room. The expansion can provides for expansion and contraction of the three circuits and also redirects the circuits west towards the experimental cave.

The transfer line extends westward approximately 60 feet to above the cave where it connects to a vertical expansion can which drops the circuits down into the cave and redirects them in the northward beam-line direction. Once in the test cave, the transfer line connects to two isolation boxes. One is for the CC1 cryomodule and the second is for the CC2 cryomodule. Each isolation box connects to their respective CC cryostats using u-tubes which provide positive isolation from the rest of the cryogenic system should the cryostat require isolation from the rest of the cryogenic system. This is necessary to minimize thermal-cycling of the CM1 cryomodule system. Only the CC2 system will be connected for the initial phase of commissioning.

Downstream of the two isolation boxes is a cooldown box with a Joule-Thomson valve to redirect the helium flow back to the refrigerator. The cooldown box also contains a helium "mixing chamber" for controlling the helium temperature supplying the CM1 cryomodule system. The mixing chamber consists of a baffled vessel which mixes the helium transfer line supply from the refrigerator with a regulated room temperature helium gas supply to control the outlet temperature to the cryomodule system during initial cooldown of the cryomodule. The cooldown box also contains control valves in order to redirect the nitrogen flow out to an external 15 kW vaporizer unit for fine temperature control of the nitrogen flow leading to the cryomodule system also during initial cooldown.

The transfer line extends out of the cooldown box north to the cryomodule distribution system. Currently this transfer line section is isolated for the initial commissioning tests and will be connected in the near future. The cryomodule distribution system (CDS) consists of a feed box/feed cap which interfaces to the CM1 and an end cap that redirects the flows back. The CDS components were designed and built by Cryotherm GmbH & Co., Germany and are similar to those installed at CMTB, DESY [2].

**2K Systems**

In order to operate the cavities at 2 K, an ambient temperature pumping cycle is used consisting of a Kinney® roots blower (Model KMBD 10,000) and a Kinney® liquid ring pump (Model KLRC 2100) located on a common skid in the Northeast corner of the New Muon Lab building. The vacuum skid was originally assembled at Thomas Jefferson National Accelerator Facility. It was modified for sub-atmospheric operation by reinforcing the static and dynamic seals. An identical vacuum pump skid is in service as part of the MDB cryogenic system which has had several years of trouble-free operation. The vacuum skid is connected to the cryogenic load through a long pumping header which extends out of the test cave and wraps around the building. The cryogenic load consists of the CC1 (in the future), and CC2 cavities as well as the Cryomodule cavity system. The pumping header will typically operate at a pressure of 7 torr with each cavity system regulating its own pressure using individual control valves to roughly a 12 torr level.

The vacuum pumping header consists of several vacuum insulated sections that run along the west wall. These sections are vacuum insulated in order to prevent condensation from dripping on the cave enclosure. Prior to reaching the vacuum pump skid, the pumping line converts to a non-vacuum insulated line allowing the flow to warm up before entering the vacuum pump.

In order to support 2 K operations and to improve reliability at the New Muon Lab facility, a purifier compressor has been installed in the pump room next to the vacuum skid. An existing oil flooded 150 horsepower screw helium compressor manufactured by Frick® was refurbished and modified for this purpose. A new oil removal system was designed and built for this system as well. The maximum capacity of the Frick® compressor is estimated to be 13 g/s of helium at 1 psig inlet pressure. The inlet pressure is regulated using a discharge to suction bypass valve to maintain pressure at 1 psig.

This compressor takes the suction return gas from the vacuum pump system, compresses the gas and puts it through the purifier/adsorber system. The reasoning behind this is that due to extended sub-atmospheric operation of the 2K system there is the possibility of air migration into the system. Although the system has been designed for sub-atmospheric operation with close attention paid to seal integrity of sub-atmospheric components, the Frick® compressor and adsorber system will act as a guard in case contamination enters the system. This is warranted since the system is expected to run continuously for many months at a time.

**Controls**

The main control system at New Muon Lab consists of the Siemens Advanced Process Automation and Control System (APACS+$^{TM}$) which allows for re-configurable logic and loop control, as well as Input/Output (I/O) modules capable of handling a variety of signals. The same system has been successfully used for many years to control other cryogenic systems at Fermilab [3]. The control of the localized equipment such as the main compressors, expansion engines, vacuum pump and purifier compressor are done using localized, self-contained, PLC based controls (*Direct*Logic205 PLCs by KOYO$^®$) which communicate directly with the APACS system using a fiber optic line. The localized PLC's interface with the equipments motor controller and manage the machine interlocks. The start/stop/reset features, the remote/local control as well as a limited amount of input and output channels are also managed by this PLC. Locally, a touch panel display allows for manipulation and control of these systems and parameters.

The top layer human machine interface used for the New Muon Lab Cryogenic system is iFIX from GE Fanuc Intelligent Platforms. The iFIX system is a graphical interface between the APACS+$^{TM}$ system and the end user which uses graphical tools to display the cryogenic process. Control of the system can also be done using these tools by simply clicking on graphical components and manipulating the output. The displays are created using the iFIX graphical data display builder. The iFIX system also supports alarm handling, data archiving and plotting packages.

Many of the I/O devices are also sent to the Fermilab ACNET control system from APACS+$^{TM}$ using an OPC server. This flexibility gives experimenters access to data from various systems in one platform for ease of plotting and data management. Remote alarm capabilities are performed using WIN-911$^®$ alarm notification software which provides real-time automatic notification via a phone line, pager or email. Many alarms are also broadcast on the ACNET system, which are viewed in several control rooms at Fermilab.

**CRYOGENIC REQUIREMENTS**

The theoretical capacity of a single Tevatron satellite refrigerator is either: 625 W at 4.5 K in refrigerator mode or 4.17 g/s of liquid helium at atmospheric pressure in liquefaction mode. Capacity measurements of a Tevatron satellite refrigerator were conducted at the Fermilab Cryogenic Test Facility (CTF) and agreed well with the theoretical numbers, achieving a sustainable capacity of ~3.7 g/s and ~50 W at 4.5 K [4]. Further tests conducted at New Muon Lab in the fall of 2007 as part of the South refrigerator commissioning indicated that the maximum sustainable capacity in a liquefier mode was ~2.8 g/s which is significantly lower than was measured earlier at CTF. This under capacity is attributed to insufficient time to tune the refrigerator operational loops and parameters. It is believed that higher liquefier capacities approaching those of the CTF tests are possible given enough testing time.

To understand the operational boundaries of the New Muon Lab cryogenic system, the estimated required capacity was superimposed on a plot of the theoretical and measured refrigerator capacity for two refrigerators as shown in FIGURE 1. The results indicate that operating CC2 and CM1 together is possible using one satellite refrigerator. Furthermore, with the addition of the second refrigerator it should be possible to operate CC1, CC2, CM1 and CM2 as well. Note that these results assume the refrigeration system operating at peak performance. Any degradation in expander or heat exchanger performance can reduce the refrigerator capacity significantly. In order to address these concerns, it is planned to add a new cryogenic plant to supply the cryomodule system while using the existing Tevatron refrigerator system to support the CC1/CC2 system.

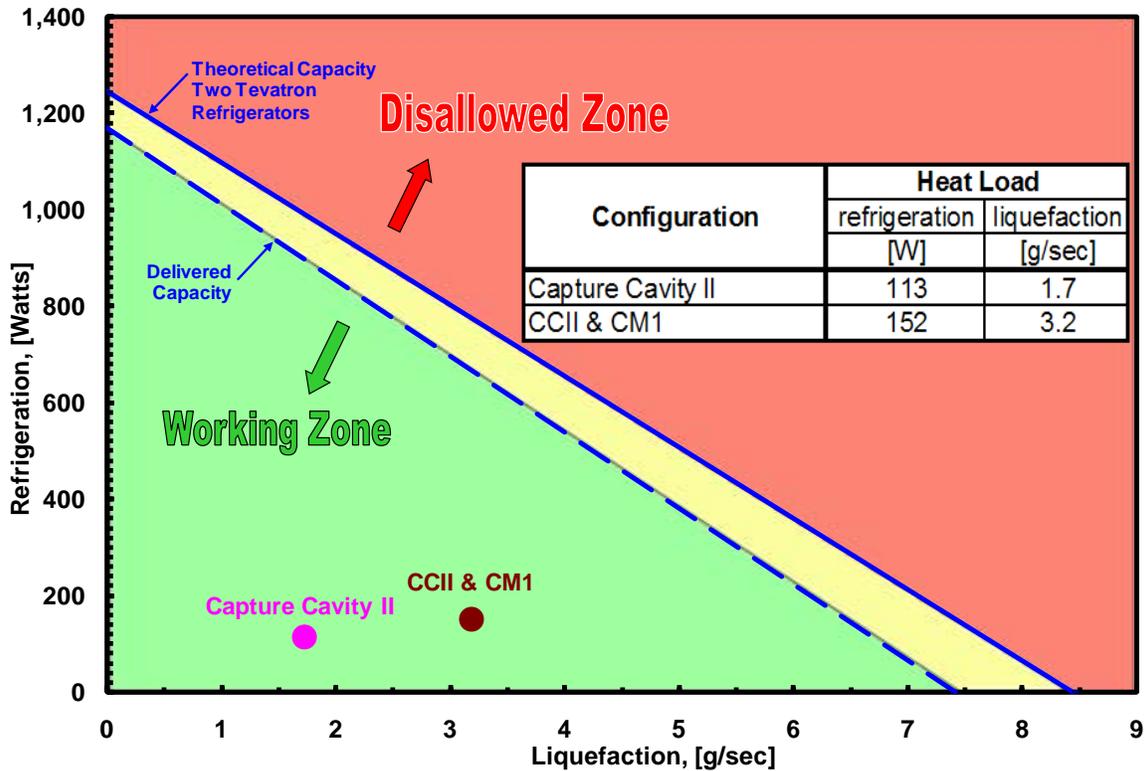

**FIGURE 1.** Refrigeration versus liquefaction capacity for the NML North and South Refrigerators.

## INITIAL COMMISSIONING

Over the last year several key components of the New Muon Lab cryogenic system have been successfully commissioned. These include the purifier compressor as well as the vacuum pump system. The standalone South refrigerator was also previously commissioned by locally producing liquid in the 500 liter helium dewar. The next phase of commissioning, Phase 1a, will include cooling down the recently installed cryogenic distribution system leading to the test cave. The test load will consist of the CC2 cryostat which has been relocated from the MDB test facility where it has operated for several years, see FIGURE 2. This phase of commissioning will allow for cryogenic and RF system performance verification prior to full CM1 operation. In order to test the RF system, a shield block wall was installed at the South end of the test cave to allow full CC2 RF and cryogenic testing in this area while still allowing for installation of the CM1 and related equipment in the other portions of the test cave.

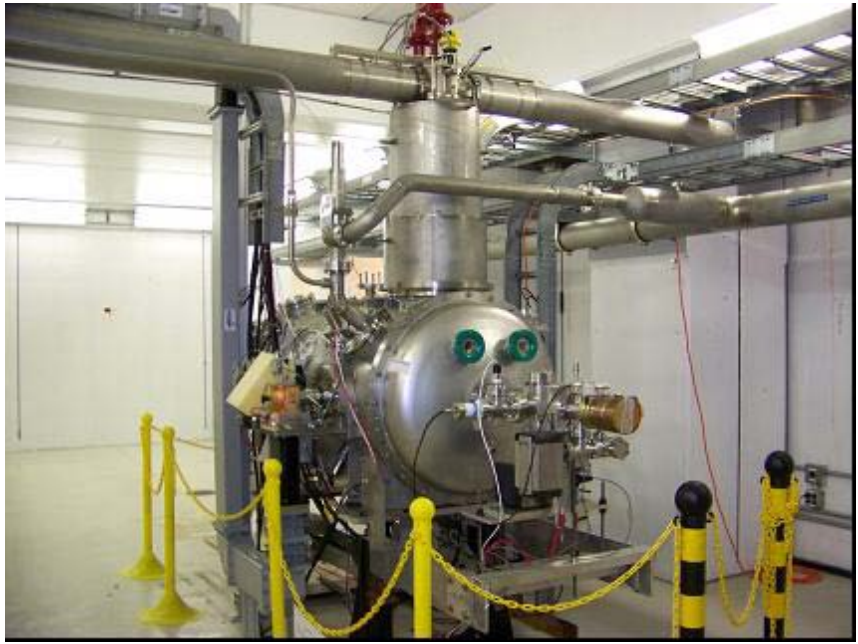

**FIGURE 2.** The Capture Cavity II 9-Cell SRF Cryomodule at New Muon Lab.

Upon completion of the Phase 1a commissioning with CC2, the shield block wall will be removed and the supply transfer line leading to the cryomodule feed box will be installed, completing the cryogenic circuit to CM1. Phase 1 testing would then involve operating the CC2 cavity with CM1. These tests would include cryogenic commissioning as well as full RF testing.

Subsequent facility development phases involve testing of an additional single cavity cryomodule, CC1. As previously mentioned, to support the additional heat load of the CC1+CC2+CM1 system tests, a second Tevatron style North refrigerator will be installed. There are also plans to extend the cave tunnel portion north to allow for installation of two ILC RF units. This addition will require the installation of a new cryoplant to supply the cryomodule system while using the existing Tevatron style North and South refrigerator system to support the CC1 and CC2 system separately.

**CONCLUSIONS**

The New Muon Lab cryogenic system is nearing completion with the successful commissioning of several subsystems including the South refrigerator, vacuum pump and compressor systems. The commissioning of the cryogenic distribution system leading to the test cave using the CC2 cryomodule will soon be underway followed by full RF commissioning of the system. Work continues on the cryogenic integration of the first 1.3 GHz Type III plus eight cavity cryomodule (CM1) using a new feed box/feed cap and end cap designed and built by Cryotherm GmbH & Co. Commissioning of the CC2 cryomodule and CM1 are expected in the fall of this year. Future plans include a building extension to support two RF units as well as a new cryogenic plant to handle the additional heat loads.

## ACKNOWLEDGEMENTS

Fermilab is operated by the Fermi Research Alliance, LLC under Contract No. DE-AC02-07CH11359. The authors wish to recognize the dedication and skills of the Accelerator Cryogenic Department technical personnel involved in the design and installation of the cryogenic equipment.